\documentclass[11pt,twoside]{article}

\usepackage{asp2006}
\usepackage{epsf}
\usepackage{epsfig}
\usepackage{lscape}
\usepackage{subfigure}

\def\kms{$\rm km\;s^{-1}$}

\def\hb{H$\beta$}

\def\oiii{[O~{\small III}]}

\begin{document}
\title{VLT/VIMOS Integral field kinematics of the Giant Low Surface
Brightness galaxy ESO 323-G064} 
\author{Coccato,
L.$^1$, Swaters, R.$^2$, Rubin, V.C. $^3$, D'Odorico, S. $^4$, \&\
McGaugh, S.$^2$} 
\affil{$^1$ MPE-Garching;
Germany; $^2$ University of Maryland, USA; $^3$ Dept. of Terrestrial
Magnetism, CIW, USA; $^4$ ESO-Garching, Germany } 

\begin{abstract} 

We studied the bulge and the disk kinematic of the giant Low Surface
Brightness (LSB) galaxy ESO 323-G064 in order to investigate its
dynamics and its Dark Matter (DM) content. We observed the galaxy with
the integral field spectroscopy (VLT/VIMOS, in IFU
configuration). Results for the gaseous kinematics (bulge and disk)
and stellar kinematics (bulge) are presented, together with a Jeans
model for the stellar bulge kinematics.

\end{abstract}


\section{The giant low surface brightness galaxy ESO 323-G064}   

\noindent{\bf Introduction} LSB galaxies play a significant role in
our understanding the bright and DM distribution in the universe. Like
high surface brightness galaxies, LSB galaxies also span a range in
properties such as size, mass, and bulge size. For example, Bothun et
al. (1990) discovered a class of giant LSB galaxies, and Beijersbergen
et al. (1999) studied a sample of bulge-dominated LSB galaxies. To
date, very little is known of the properties of this new class of
objects. To fill this gap, Swaters \& Rubin (in preparation) have
recently completed a survey of the kinematics of a sample of bulge
dominated LSB galaxies to study their DM content and
distribution. Here we show the first results obtained for the giant
LSB galaxy ESO 323-G064 (at a distance of 198 Mpc), selected from that
project.

\noindent{\bf Observations} Observations of ESO 323-G042 had been
carried out with VIMOS equipped with Integral Field Unit at the Very
Large Telescope (Chile). The high resolution blue grism was used, with
a spatial resolution 0.67$\times$0.67/fiber, a spectral resolution of
$\approx$75 \kms, covering the spectral range 4120--5460 \AA. The
total exposure time is 4 hours.

\noindent{\bf Gaseous kinematics} The bulge ($R<5''$) is characterized
by complex emission-line structure: the emission lines show a
multi-peaked profile (Figure 1A). The disk ($13''<R<30''$) of the
galaxy shows more regular rotation (Figure 1B), with an amplitude of
248$\pm$6 \kms\ (Figure 1C).


\begin{figure}[t]
\centering
  \begin{minipage}[c]{0.28\textwidth}
    \hspace{-2.5cm}
  \epsfig{file=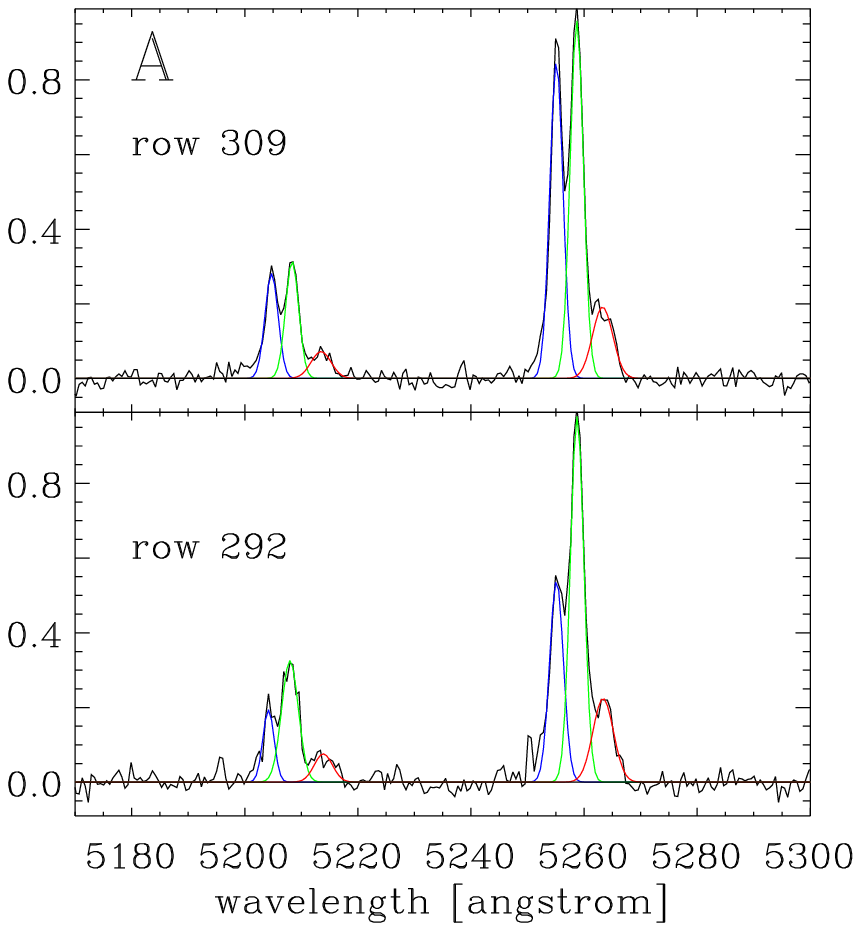,clip=,width=4.8cm}
  \end{minipage}
  \begin{minipage}[c]{0.28\textwidth}
    \hspace{-1.cm}
  \epsfig{file=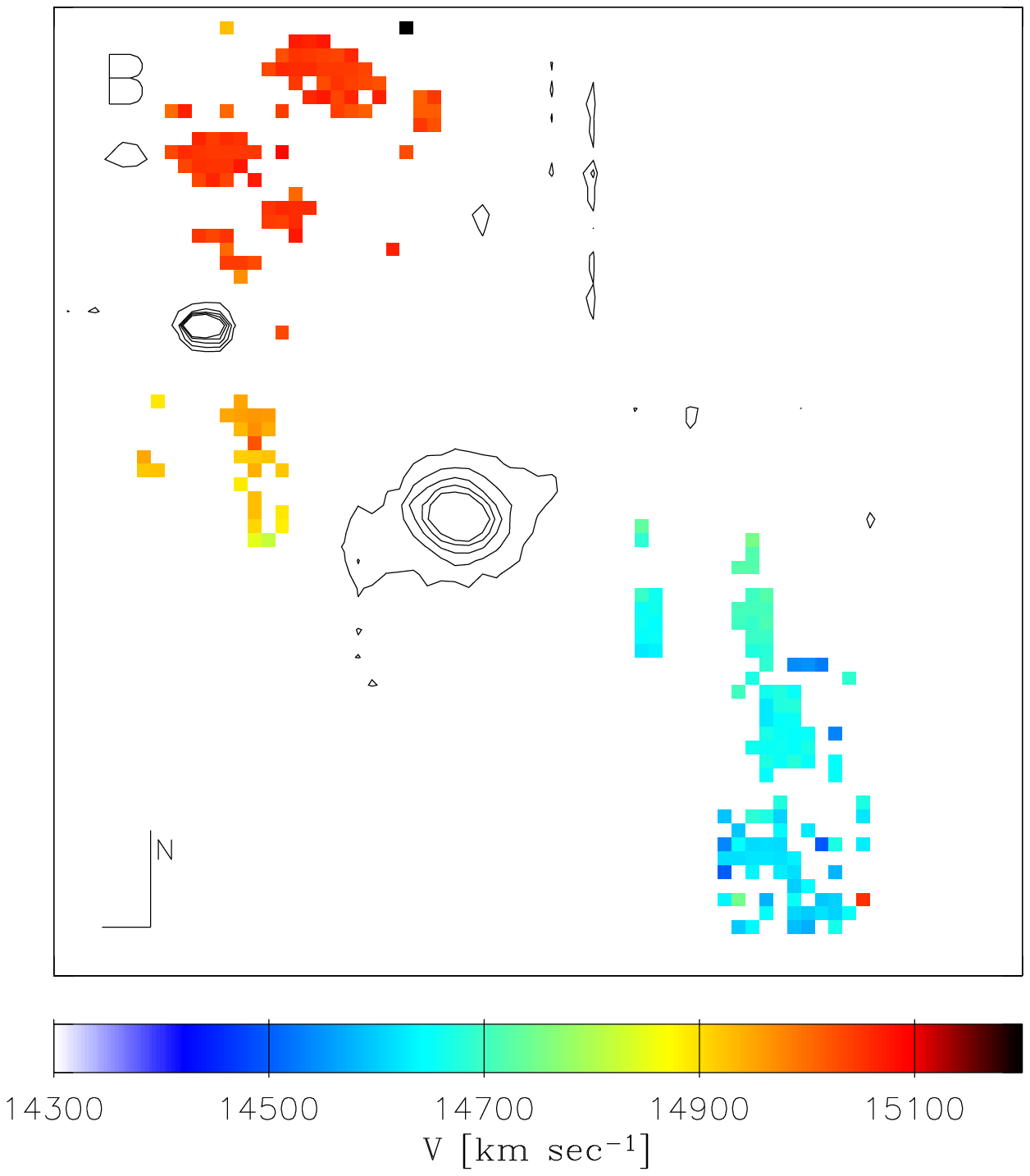,clip=,width=4.6cm}
  \end{minipage}
  \begin{minipage}[c]{0.28\textwidth}
    \hspace{2.5cm}
    \vbox{
      \vspace{-1.3cm}
      \hspace{0.5cm}
      \epsfig{file=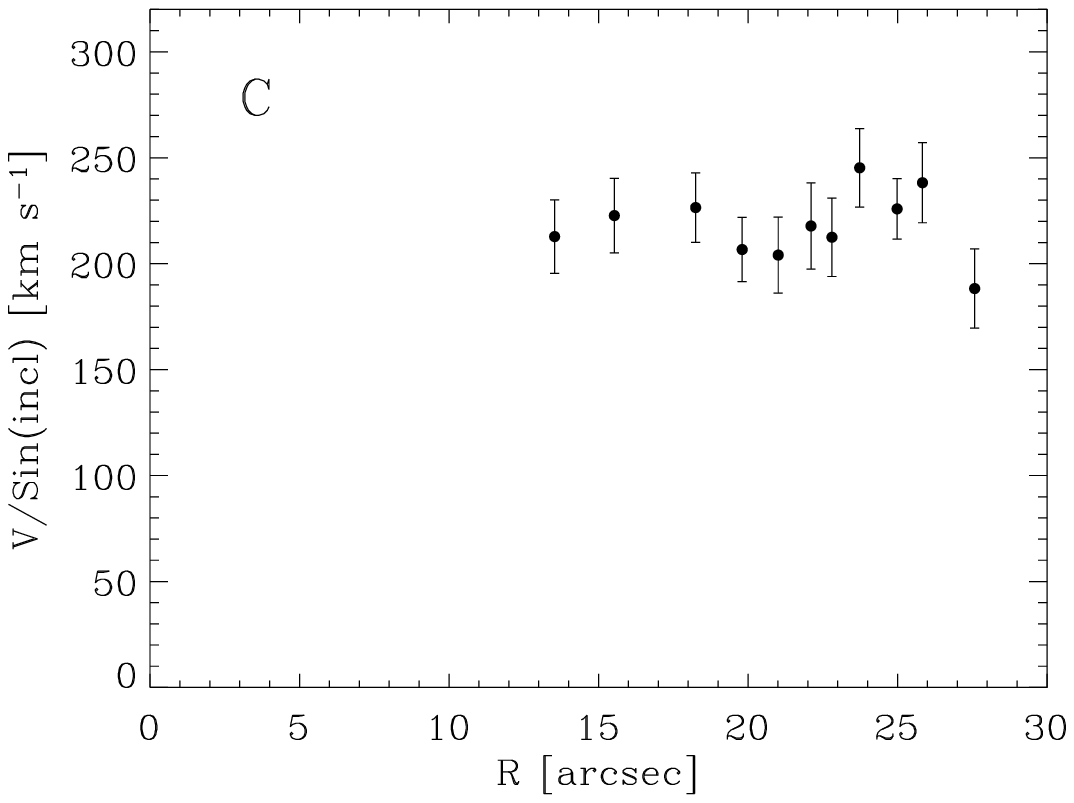,clip=,width=5.5cm}
    }
  \end{minipage}
%
  \begin{minipage}[c]{0.32\textwidth}
    \hspace{-1.50cm}   \epsfig{file=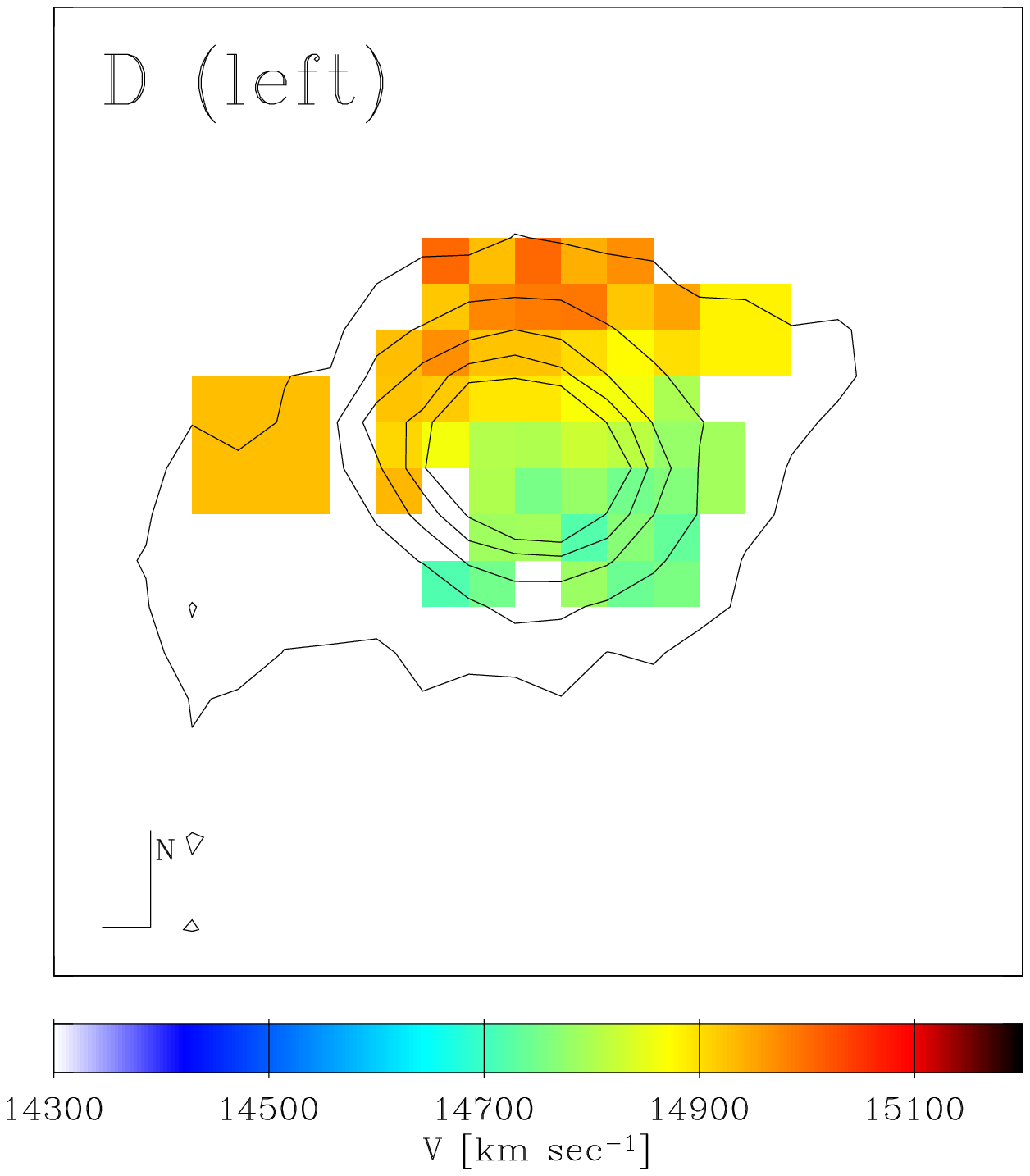,clip=,width=4.5cm}
  \end{minipage}
  \begin{minipage}[c]{0.32\textwidth}
    \hspace{-1.450cm}   \epsfig{file=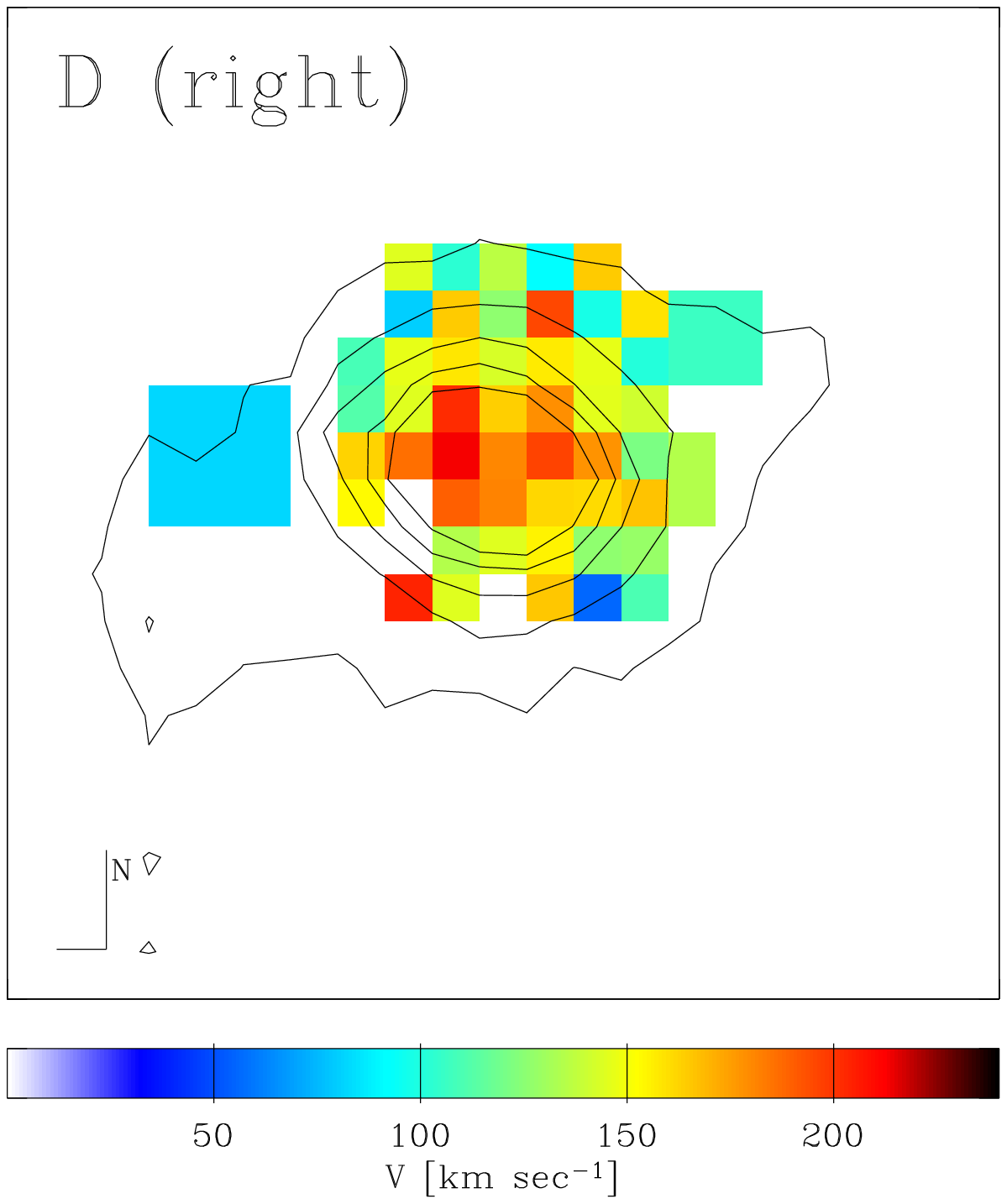,clip=,width=4.5cm}
  \end{minipage}
  \begin{minipage}[c]{0.32\textwidth}
   \epsfig{file=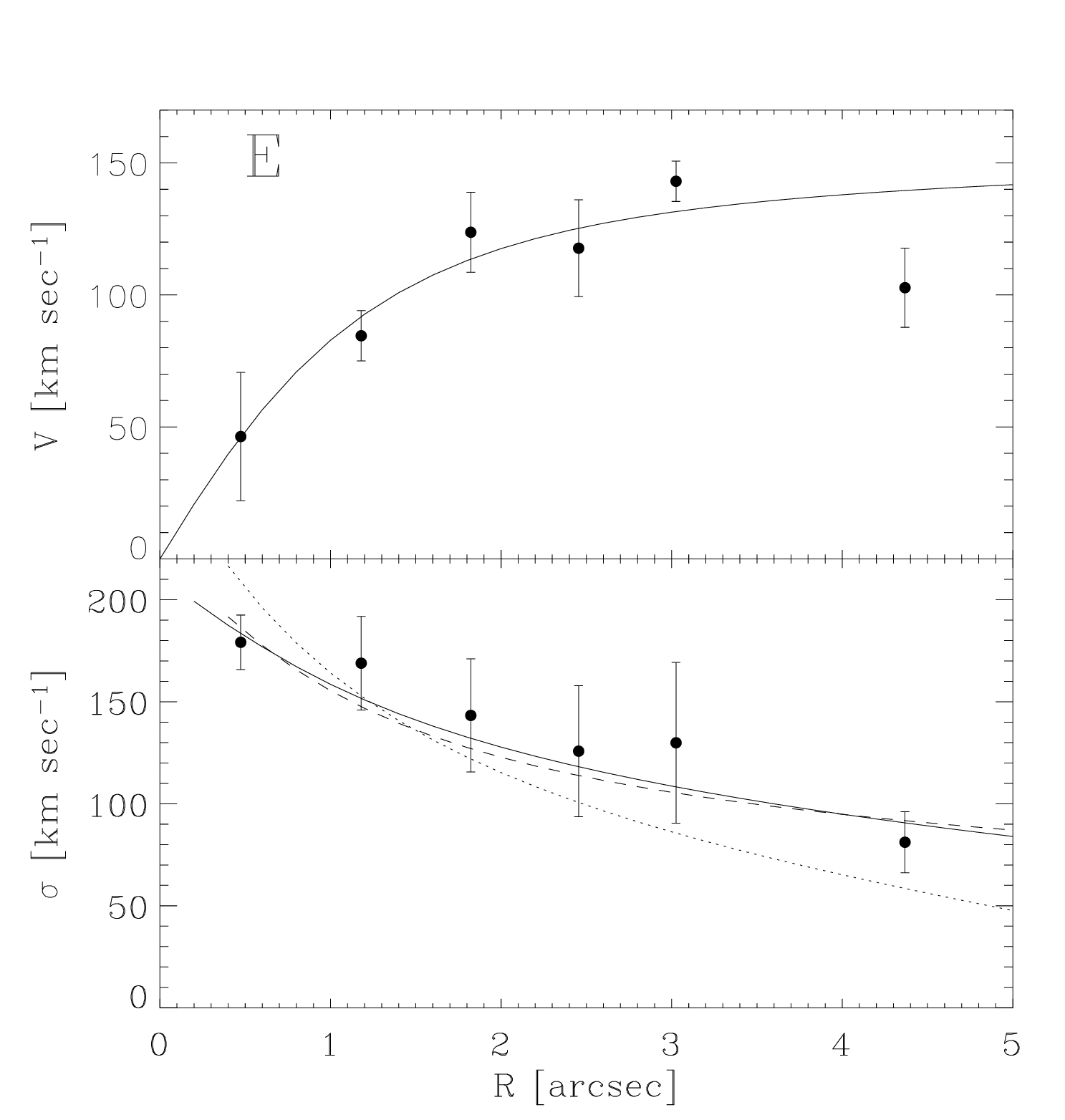,clip=,width=5.4cm}  
  \end{minipage}
\caption{A: example of redshifted \oiii\ emission lines in the bulge
with multiple peaks. B: Velocity field of the \hb\ emission line in
the disk. Field of view is $46''\times46''$. C: rotation curve
extracted from B. D: stellar velocity (left) and velocity dispersion
(right) of the bulge Field of view is $14''\times14''$. E: stellar
rotation curve (top) and velocity dispersion profile (bottom) of the
bulge with Jeans models superposed: self consistent case (dotted
line), NFW (continuous line) and pseudo-isothermal halo (dashed line).}
\end{figure}

\noindent{\bf Stellar kinematics} The stellar velocity and velocity
dispersion have been measured in the bulge $5''$, unveiling a regular
rotation with an observed amplitude of 140 \kms\ and a central
velocity dispersion of $\sigma=180$ \kms\ (Figure 1D).  The stellar
kinematics have been modeled with spherical isotropic Jeans models,
exploring the self consistent and the DM scenarios, Navarro, Frenk and
White (NFW, 1990) and pseudo isothermal halo. The limited radial
extent of the stellar kinematics did not allow us to extend the models
to the disk component and to consider more complicated scenarios
(i.e., anisotropy or bars). Our relatively simple Jeans modeling shows
that dark matter is needed in the central $5''$ to explain the bulge
kinematics, but we are not able to disentangle between different DM
scenarios (Figure 1E). In particular, the total mass of the bulge
(within $5''$) is $(7.4 \pm 3.2)\cdot 10^{10}$ M$_{\odot}$ (DM fraction 55\%)
and $(7.1 \pm 3.6) \cdot 10^{10}$ M$_{\odot}$ (DM fraction 42\%) according to
the NWF and pseudo isothermal scenarios, respectively.

\bigskip
\noindent{\bf Bibliography}

Bothun, G. D., Schombert, J. M., Impey, C. D. \&\ Schneider, S.,
E. 1990, ApJ, 360, 427

Beijersbergen, M., de Blok, W. J. G.,\&\ van der Hulst, J. M. 1999,
A\&A, 351, 903

Navarro, J. F., Frenk, C. S., White, S. D. M. 1997, ApJ, 490, 493







\end{document}